\documentclass[11pt]{article}
\usepackage[english]{babel}
\usepackage{amsmath,amsthm}
\usepackage{amsfonts}
\usepackage[margin=1.25in]{geometry}

\theoremstyle{definition}

\theoremstyle{remark}

\numberwithin{equation}{section}
\begin{document}
\title{Effect of bubble deformability on \\the vertical channel bubbly flow}
\author{Sadegh Dabiri$^1$, Jiacai Lu$^2$, Gretar Tryggvason$^1$ \\
\\\vspace{6pt} $^1$University of Notre Dame, Notre Dame, IN 46556, USA \\ $^2$Worsecter Polytechnic Institute, Worcester, MA 01609, USA}
\maketitle
\begin{abstract}
This article describes the fluid dynamics video: ``Effect of bubble deformability on the vertical channel bubbly flow.'' The effect of bubble deformability on the flow rate of bubbly upflow in a turbulent vertical channel is examined using direct numerical simulations. A series of simulations with bubbles of decreasing deformability reveals a sharp transition from a flow with deformable bubbles uniformly distributed in the middle of the channel to a flow with nearly spherical bubbles with a wall-peak bubble distribution and a much lower flow rate.\end{abstract}
\section{Introduction}
A series of numerical simulations of an upward bubbly flow with a void fraction of 3\% between two vertical walls using direct numerical simulation with a front tracking method is performed. A rectangular computational domain with two no-slip vertical wall and periodic boundary conditions in streamwise and spanwise directions is used. The effect of bubble deformability is isolated through controlling the surface tension while keeping other flow parameters fixed. The flow is driven upward by a specified pressure gradient. The dimensionless numbers in the problem are Reynolds number, $Re=\frac{\rho uw}{\mu}$, Eotvos number, $Eo=\frac{\rho g^2 d}{\sigma}$, and density and viscosity ratios. Here, $\rho,\mu,\sigma,d,w,u$, and $g$ are the liquid density and viscosity, surface tension coefficient, bubble diameter, channel width, average vertical velocity in the channel, and acceleration due to gravity, respectively.
The pressure gradient and Eotvos number are specified \emph{a priori} and the Reynolds number is calculated from the average velocity \emph{a posteriori}. Eotvos number is decreased from 4.5 to 0.9. The decrease in the deformability of bubbles changes the lateral lift force on the bubbles. Therefore, the distribution of the bubbles changes from a uniform distribution in the middle of the channel to a wall-peak distribution with only a few bubbles in the center. The reduction of the number of the bubbles in the center increases the average density there which leads to a reduction in flow rate since the driving pressure gradient is kept fixed. As a result, the Reynolds number drops from 3750 to 1700.

\end{document}